\documentclass[twocolumn,10pt]{asme2ej}

\usepackage{graphicx} %% for loading jpg figures
\usepackage{xcolor}

\newcommand{\eg}{{\em e.g.,}}
\newcommand{\ie}{{\em i.e.,}}

\newcommand{\etal}{{\em et~al.}}
\newcommand{\red}[1]{\textcolor{black}{#1}}
\usepackage{setspace}
\usepackage{mathtools}
\usepackage{multirow}
\usepackage{booktabs}

\usepackage[ruled]{algorithm2e} % For algorithms

\SetAlFnt{\small}
\SetAlCapFnt{\small}
\SetAlCapNameFnt{\small}
\SetAlCapHSkip{0pt}

\title{Manufacturability Oriented Model Correction and Build Direction Optimization for Additive Manufacturing}
\author{Erva Ulu\thanks{Address all correspondence to this author.} , Nurcan Gecer Ulu, Walter Hsiao, Saigopal Nelaturi
    \affiliation{
	Palo Alto Research Center\\
	Palo Alto, California 94304\\
    Email: eulu@parc.com
    }	
}

\begin{document}

\maketitle    

%%%%%%%%%%%%%%%%%%%%%%%%%%%%%%%%%%%%%%%%%%%%%%%%%%%%%%%%%%%%%%%%%%%%%%
\begin{abstract}
{\it We introduce a method to analyze and modify a shape to make it manufacturable for a given additive manufacturing (AM) process. Different AM technologies, process parameters or materials introduce geometric constraints on what is manufacturable or not. Given an input 3D model and minimum printable feature size dictated by the manufacturing process characteristics and parameters, our algorithm generates a corrected geometry that is printable with the intended AM process. A key issue in model correction for manufacturability is the identification of critical features that are affected by the printing process. To address this challenge, we propose a topology aware approach to construct the allowable space for a print head to traverse during the 3D printing process. Combined with our build orientation optimization algorithm, the amount of modifications performed on the shape is kept at minimum while providing an accurate approximation of the as-manufactured part. We demonstrate our method on a variety of 3D models and validate it by 3D printing the results. 
}
\end{abstract}

\section{Introduction}

\begin{figure*}
  \centering
  \includegraphics[width = \textwidth]{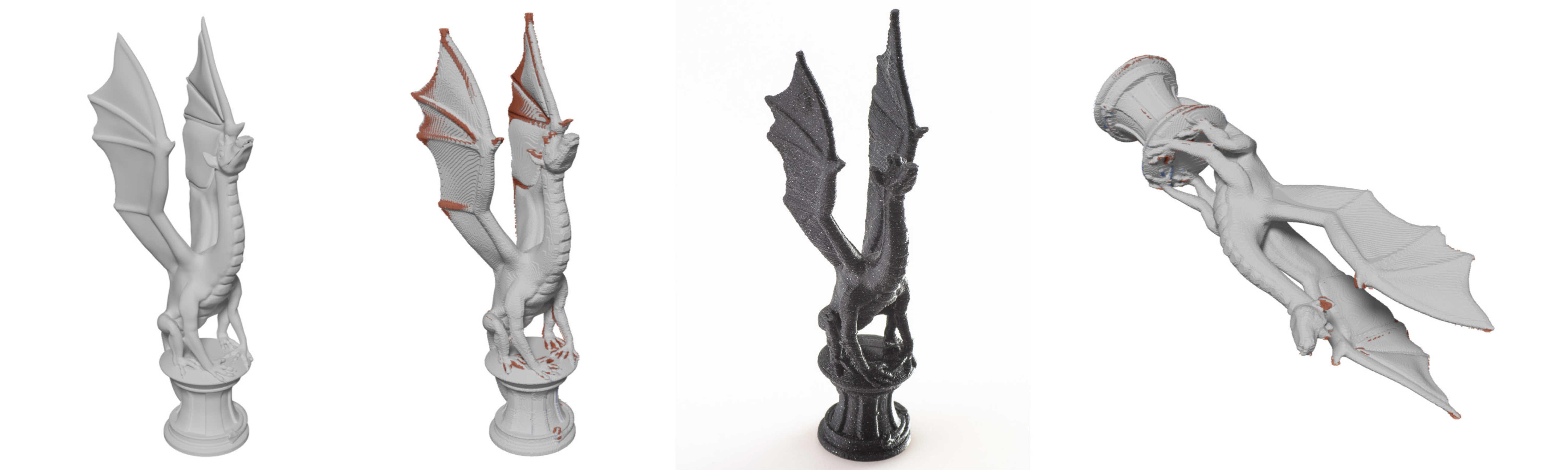}
  \caption{We present a method for correcting 3D models to make them manufacturable with the intended additive manufacturing process. Our approach analyzes the input model (gray) layer by layer and provides modifications to the shape (in the form of addition (red) or removal (blue) of material) to avoid fabrication errors due to the minimum printable feature size. Required modifications are minimized by optimizing the build direction (right).}
  \label{fig:teaser}
\end{figure*}

Recent advances in additive manufacturing technologies have triggered the development of powerful design methodologies allowing designers to create highly complex functional parts \cite{schumacher2015microstructures,martinez2016procedural,liu2018narrow,arora2018designing,ulu2019structural}. Many such methods often operate under the assumption that any designed shape can be fabricated using a 3D printer and the resulting part matches the designed shape perfectly or with negligible errors. In reality, however, the selected AM process, mechanical characteristics of the printer or the material being used may introduce significant limitations to printability of a particular design in the form of a minimum printable feature size. %constraints to the feasible design space.%
For instance, the minimum feature size might be relatively large for a consumer level fused deposition modeling (FDM) printer while an order of magnitude smaller features could be achieved with a high-end stereolithography (SLA) printer~\cite{bhushan2017overview}. Similarly, in electron beam melting (EBM) process, higher beam power leads to a larger melt pool, thereby resulting in a larger minimum printable feature size~\cite{ulu2019concurrent}. Therefore, attempting to manufacture a complex shape with very small features using a consumer level FDM printer or an EBM machine with high beam power settings could lead to poor quality parts or even complete failure of the print. In order to avoid such problems during the printing process, it is crucial to take the manufacturing constraints into account during the design stage and modify the shape accordingly.

We propose a new approach for analyzing and altering the designed geometry to make it manufacturable with the intended 3D printing approach and corresponding process parameters. Our approach takes as input (1) a 3D shape represented by its boundary surface mesh and (2) a minimum printable feature size dictated by the manufacturing process, and produces a {\it corrected} model that accurately approximates the as-manufactured part which can be 3D printed without any failures (Figure~\ref{fig:teaser}).

In analyzing and correcting a model for manufacturability, the main challenge lies in identifying the important features and determining the modifications to the design that affects the shape minimally. Our approach overcomes this challenge by constructing a {\it meso-skeleton} --the maximal area within each slice where a print head can be positioned  during the printing process-- that is topologically equivalent to the corresponding slice of the input shape. This approach allows us to thicken topologically important parts that are smaller than the minimum printable size by creating a minimal path for the print head to deposit material. We keep these modifications minimal by optimizing the build orientation of the part. Our approach operates on the slice level, simulating the layer-by-layer printing process, and thus, the resulting geometry accurately approximates the printed part revealing the downstream geometric changes in a design-to-manufacturing pipeline.  

Our main contributions are:
\begin{enumerate}
\item a novel topology aware model correction method for manufacturability,
\item a method to construct a meso-skeleton that serves as a proxy to the maximal area that the print head can traverse,
\item a build orientation selection algorithm to minimize the amount of required shape modifications.
\end{enumerate}

\section{Related Work}
Our review is comprised of studies highlighting {\it design for fabrication constraints}, {\it manufacturability analysis and model correction}, and {\it build direction selection} with an emphasis on approaches involving additive manufacturing. 

\subsection{Design for Fabrication Constraints}
A large body of work has investigated automatic techniques offering design aids in creating shapes that satisfy a variety of fabrication related constraints \cite{livesu2017from,bermano2017state}. Recent examples include partitioning and packing approaches to accommodate large prints \cite{luo2012chopper,vanek2014packmerger,chen2015dapper}, \red{advanced slicing methods to improve the print quality and build time \cite{huang2013intersection,wang2015saliency,alexa2017optimal,mao2019adaptive}}, automatic support structure design methods \cite{vanek2014clever, dumas2014bridging} as well as shape modification approaches to minimize or avoid support structures  \cite{hu2015support,xie2017support,wei2018toward}. Other shape modification approaches have been explored to meet geometric requirements such as minimum feature size or wall thickness \cite{wang2013thickening,cabiddu2017e,attene2018as}. Our approach falls under this general category in that we alter the geometry to make it manufacturable under the given minimum printable feature size limitations. Our aim, however is to provide an accurate approximation of the as-manufactured part by performing a slice based analysis to construct an allowable space for the print head to traverse during the 3D printing process.

%identify the geometric incompatibilities of 3D models 

\begin{figure}
  \centering
  \includegraphics[width = \columnwidth]{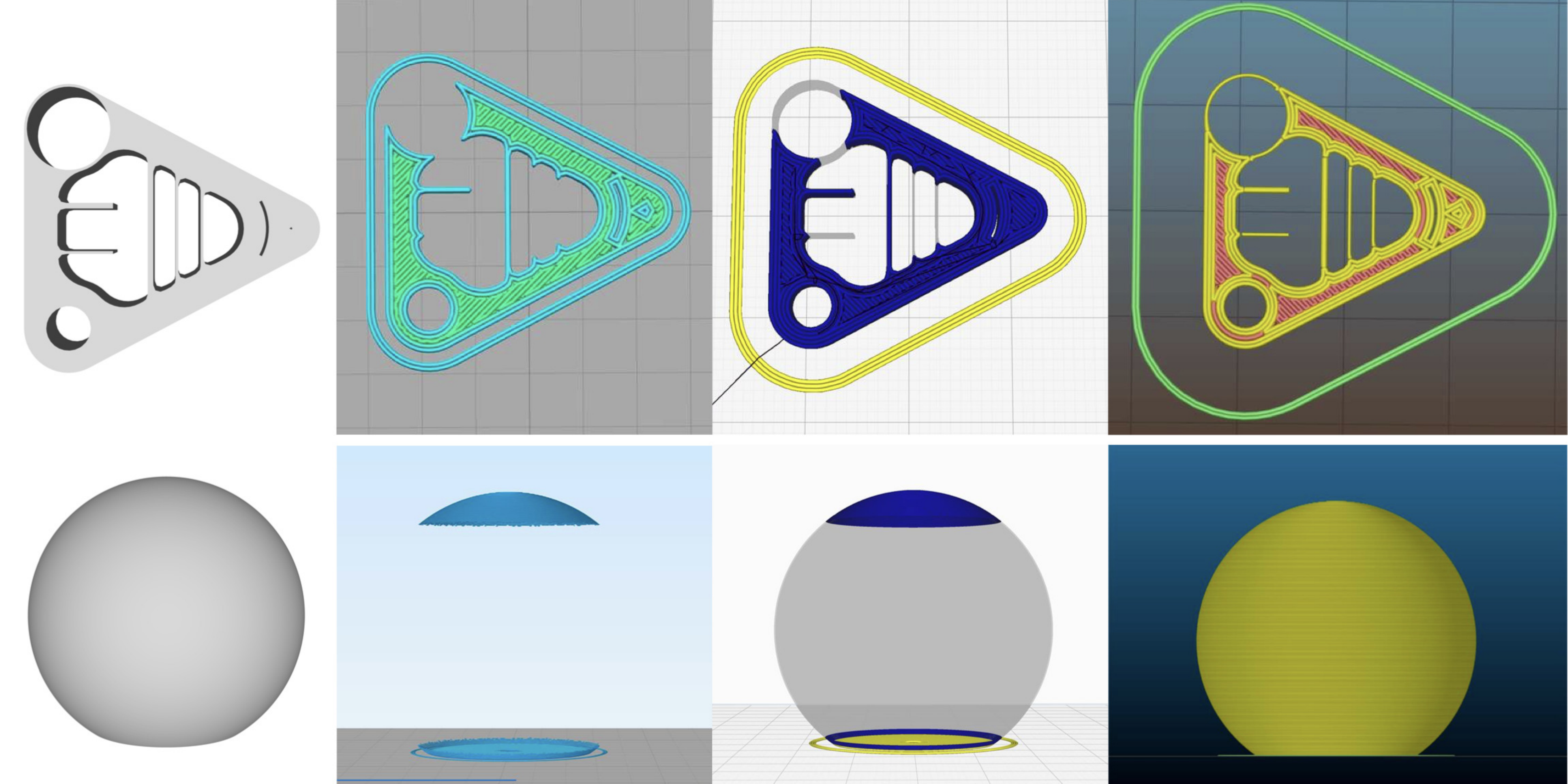}
  \caption{Toolpaths generated by some of the commonly used slicers for two example models: a test part with thin features (top) and a thin spherical shell (bottom).  Left-to-right: the input model, toolpaths generated by Simplify3D \cite{simplify3d}, Ultimaker Cura \cite{cura} and Slic3r (Prusa edition) \cite{slic3r}.}
  \label{fig:slicerExamples}
\end{figure}

\subsection{Manufacturability Analysis and Model Correction}
In the context of additive manufacturing, conventional manufacturability analysis approaches are often intended for detecting artifacts in the input surface mesh such as self-intersections, non-manifoldness and non-watertightness \cite{ju2004robust,attene2013polygon}. However, even models free of such artifacts may not be 3D printable as the size of certain regions of the object might drop below the printing resolution \cite{moylan2012proposal}. Morphological operations and distance transforms \cite{telea2011voxel}, ray casting \cite{tedia2016manufacturability} and feature recognition \cite{shi2018manufacturability} based approaches have been presented recently to identify such problematic regions. Although these approaches are demonstrated to provide valuable design feedback to user by determining compatibility of a 3D model with the printing hardware, very few solutions exist to automatically modify the shape to mitigate these manufacturability problems. 

\begin{figure*}
  \centering
  \includegraphics[width = \textwidth]{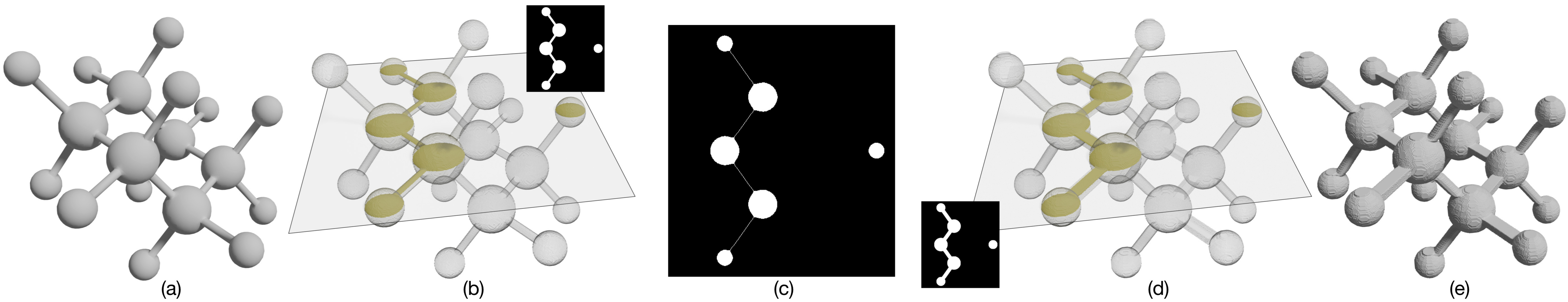}
  \caption{Given an input 3D shape (a) and a minimum printable feature size (or a structuring element size), our algorithm voxelizes the shape to generate 2D slices corresponding to layers in the printing process (b). Each layer is then analyzed to construct a meso-skeleton (c). Dilation of each meso-skeleton with the structuring element results in the corrected slice (d). The stack of corrected slices is used to generate the boundary surface of the corrected 3D model (e). Note that our approach preserves thin skeletal paths (c) that would disappear through conventional morphological opening operations.}
  \label{fig:overview}
\end{figure*}

In \cite{jaiswal2019geometric}, problematic regions are identified using 2D shape diameter function and then, a physics based mesh deformation technique followed by a post-processing step using morphological operations is used to make local corrections to improve printability of the design. In this approach, small holes and narrow intrusions are aimed to be eliminated as a part of the correction process. While the provided corrective actions improve the design significantly, it has been shown that the printability is not guaranteed. Nelaturi~\etal~\cite{nelaturi2015manufacturability} use medial axis transformation together with techniques from mathematical morphology to construct a printability map and recommend design modifications. However, as the medial axis computation is very sensitive to noise, many unwanted branches may be created, especially on high curvature regions such as corners. Therefore, a special care has to be taken to remove them. We are inspired by this approach in that we use a skeletal structure to preserve topologically and geometrically important features when correcting the shapes for manufacturability. In order to address the sensitivity problems however, we utilize thinning methodologies \cite{lam1992thinning} in binary image processing to construct our meso-skeleton.

\subsection{Build Direction Selection}
Effects of build orientation on the build time and cost \cite{alexander1998part,ahn2007fabrication},  the amount
of support material \cite{ezair2015orientation,morgan2016part}, the mechanical properties \cite{umetani2013cross,ulu2015enhancing} and the surface quality \cite{zhang2018perceptual} have been studied extensively and automatic build direction selection algorithms are proposed to minimize such directional biases. Other works focus on geometrical inaccuracies caused by the staircase artifacts inherent to layered manufacturing processes \cite{hildebrand2013orthogonal, wang2016improved}. Although driven by similar motivations, in this work, we optimize the build direction in order to minimize the amount of modifications required to make the shape manufacturable with the given manufacturing process. Moreover, our formulation is complementary to the  aforementioned methods in that they may be used together to improve manufacturability of a design while also targeting other objectives.

\section{Shape Modifications}

In a traditional additive manufacturing pipeline, the 3D model is sliced into layers and then, each layer is individually processed to generate a set of machine instructions, usually in the form of a toolpath plan. The sequence of the movements in toolpath aims to track the outer boundary of the slice as well as to fill the inner regions and possible support structures. The conventional approach in toolpath planning is to generate machine instructions such that the deposited material is restricted to the interior of the slice. For a given slice $S_i$ and a minimum printable feature $F$ represented by a circle of diameter $d$, the maximal allowable region $M_i$ that the print head can traverse can be calculated by the morphological erosion operation as
\begin{equation}
M_i = S_i \ominus F. 
\label{eq:erosion}
\end{equation}
Therefore, the printed slice geometry $P_i$ can be approximated by the morphological opening of $S_i$ by $F$
\begin{equation}
\begin{aligned}
P_i &= (S_i \ominus F) \oplus F \\
&= M_i \oplus F
\end{aligned}
\label{eq:opening}
\end{equation}
where $\ominus$ and $\oplus$ denote morphological erosion and dilation operations, respectively. However, this approach causes features that are smaller than $F$ to disappear in the created toolpath and thus, in the fabricated part. Figure~\ref{fig:slicerExamples} illustrates example cases and corresponding toolpaths generated by various slicer software commonly used for FDM process. Note that, in some cases such as the thin spherical shell model, these problems manifest themselves as complete removal of layers and therefore, result in failure of the entire print process. A number of slicers attempt to address this issue by performing corrective actions in toolpath level. For instance, Slic3r adjusts the filament feed rate to an amount below what is optimal to be able to print walls thinner than the nozzle diameter (Figure~\ref{fig:slicerExamples}, right column), or  IceSL \cite{lefebvre2013icesl} provides an option to thicken the critical regions by over-depositing material. However, modifications that are made in the toolpath level usually target a particular AM process (\eg~ the above two examples target FDM process) and they may create ambiguities between what is designed and what is manufactured. Moreover, most commercial AM machines (such as Stratasys Objet or Dimension Elite) only support toolpath planning through their proprietary software that may not perform any corrective actions. Although our approach is similar to above mentioned tools in that it operates on slice level, we aim to correct the input model during the design stage, such that the design itself approximates the as-manufactured part as closely as possible and can be printed using the intended AM process.

\begin{figure*}
  \centering
  \includegraphics[width = \textwidth]{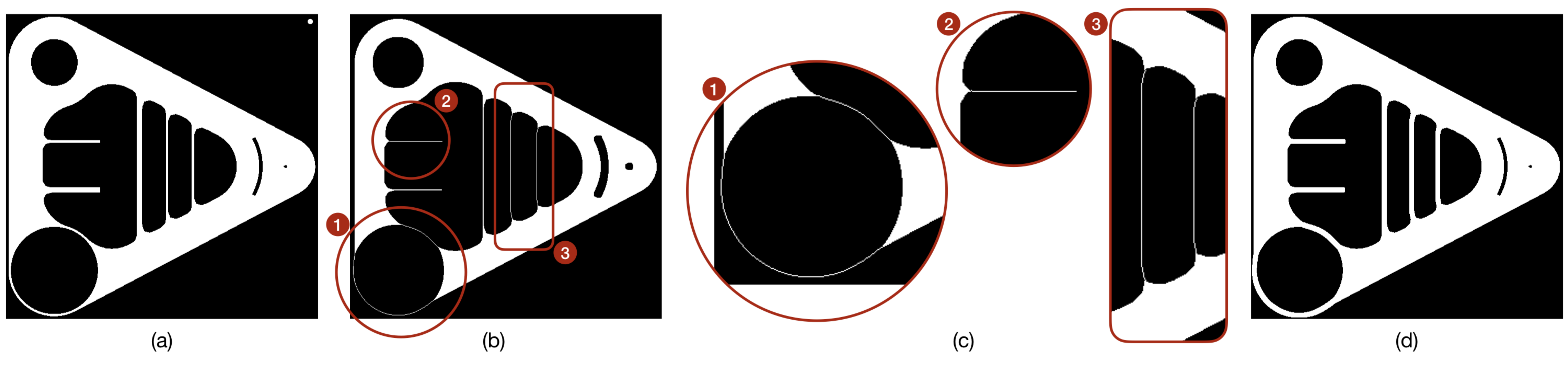}
  \caption{An example slice of the test part (a) and corresponding meso-skeleton created by our algorithm (b). Notice that skeletal paths are created for topologically important features such as (1) and (3) as well as geometrically distinct regions such as the protrusion (2) in (c). In the resulting corrected model (d), all of these problematic thin features are automatically thickened. The minimum printable feature (or the structuring element) is shown in upper right corner in (a).}
  \label{fig:distinctFeatures}
\end{figure*}

\begin{algorithm}
 \SetAlgoLined
 \SetKwInOut{Input}{Input}\SetKwInOut{Output}{Output}
 \Input{$\mathcal{B}$ , $F$}
 \Output{$\hat{\mathcal{B}}$}
 $\mathcal{V} \gets $Voxelize $\mathcal{B}$\;
 $\hat{\mathcal{V}} \gets \emptyset$\;
 \ForEach{Slice $S_i \in \mathcal{V}$}{
 $\hat{M}_i \gets$ Compute meso-skeleton\;
 $\hat{S}_i \gets \hat{M}_i \oplus F$\;
 Append $\hat{S}_i$ to $\hat{\mathcal{V}}$\;
 }
 $\hat{\mathcal{B}} \gets$ Compute boundary surface of $\hat{\mathcal{V}}$\;
 \caption{Our shape modification algorithm}
 \label{alg:ourAlgorithm}
\end{algorithm}

Figure~\ref{fig:overview} illustrates the overview of our model correction approach. We start by voxelizing the input 3D shape $\mathcal{B}$ such that each slice of voxels in z-direction represents a layer in the 3D printing process. Then, for each slice $S_i$, we construct a {\it meso-skeleton}, $\hat{M}_i$. Similar to \red{$M_i$} in \eqref{eq:erosion}, $\hat{M}_i$ defines the allowable region that the print head can traverse while printing. However, it contains additional skeletal paths for the features that are smaller than $F$ in size. Therefore, sweeping the structuring element $F$ over the meso-skeleton creates the corrected slice where the smallest feature is guaranteed to be equal or larger than $F$ in size. Finally, stacking all the corrected slices together results in the corrected 3D model. Algorithm~\ref{alg:ourAlgorithm} summarizes our method. Here, $\mathcal{V}$ is the voxelized model, $\hat{\mathcal{V}}$ is the corrected voxelization that is composed of corrected slices $\hat{S}_i$ and $\hat{\mathcal{B}}$ represents the corrected surface mesh that is ready to be printed. \red{Note that the corrected slices $\hat{S}_i$ can also be used directly to print the modified geometry using AM processes that support image based representation of layers, such as digital light processing (DLP). For other AM methods, perimeter polygons may be extracted from $\hat{S}_i$ in order to generate the toolpath required by the printer. For the sake of generality, we construct the corrected model $\hat{\mathcal{B}}$ in the form of a boundary surface mesh that is widely supported by various AM processes. In addition, $\hat{\mathcal{B}}$ provides an accurate approximation of the as-manufactured part to the user.}

\begin{algorithm}
 \SetAlgoLined
 \SetKwInOut{Input}{Input}\SetKwInOut{Output}{Output}
 \Input{$S_i$ , $F$}
 \Output{$\hat{M}_i$}
 $\red{M_i} \gets S_i \ominus F$\;
  $^{p}\hat{M}_i \gets \emptyset$\;
 $\hat{M}_i \gets S_i$\;
 $j \gets 0$\;
 \While{$\hat{M}_i \neq {}^{p}\hat{M}_i$ }{
 $^{p}\hat{M}_i \gets \hat{M}_i$\;
 Apply thinning to $\hat{M}_i$ \;
 \If{$j < d/2$}
 {
 Remove spur pixels in $\hat{M}_i $\;
 }
  $\hat{M}_i \gets \hat{M}_i \lor \red{M_i}$\;
  $j \gets j+1$\;
 }
 \caption{Our meso-skeleton computation algorithm}
 \label{alg:ourMesoSkeletonAlgorithm}
\end{algorithm}

\subsection{Meso-Skeleton Computation}
We compute the meso-skeleton of a slice by performing a topology preserving thinning operation. Given an input slice $S_i$ represented by a binary image where white pixels represent internal points and black pixels are background, we remove contour pixels that are not essential to its topology. Iterative removal of such pixels result in a {\it thinned image} that constitutes our meso-skeleton.    

Algorithm~\ref{alg:ourMesoSkeletonAlgorithm} summarizes our approach to compute the meso-skeleton in detail. We start by computing the conventional erosion \red{$M_i$} of the input slice. We use the erosion as an upper bound in the thinning process, \ie~ the meso-skeleton can not be shrunk beyond \red{$M_i$}. We ensure this by performing thinning operations on the input image only until the union of the thinned image $\hat{M}_i$ with \red{$M_i$} does not change. We employ the two-subiteration thinning algorithm in \cite{guo1989parallel}. Here, the deletion or retention of a pixel $p$ is determined by the configuration of the pixels in its local 8-connected neighborhood, $N(p)$. Under this scheme, a contour pixel $p$ is deleted if

\begin{equation}
\begin{aligned}
&(a)\qquad &&X_H(p) = 1, \\
&(b)\qquad &&2 \leq min(n_1(p), n_2(p)) \leq 3,\\
& && n_1(p) = \sum_{k=1}^{4} x_{2k-1} \lor x_{2k}, \\
& && n_2(p) = \sum_{k=1}^{4} x_{2k} \lor x_{2k+1}, \\
&(c.1)\qquad &&(x_2 \lor x_3 \lor \lnot x_8) \land x_1 = 0,\\
&(c.2)\qquad &&(x_6 \lor x_7 \lor \lnot x_4) \land x_5 = 0,\\
\end{aligned}
\label{eq:thinning}
\end{equation}
where $X_H(p)$ is the crossing number as defined in \cite{hilitch1969linear,lam1992thinning} and $x_k$ are the values of pixels in $N(p)$ with $k=1, \ldots ,8$, starting with the east neighbor and numbered in counter-clockwise order. In the first subiteration, pixels satisfying (a), (b), (c.1) and in the second subiteration, pixels satisfying (a), (b), (c.2) are removed. 

\begin{figure}
  \centering
  \includegraphics[width = 0.9\columnwidth]{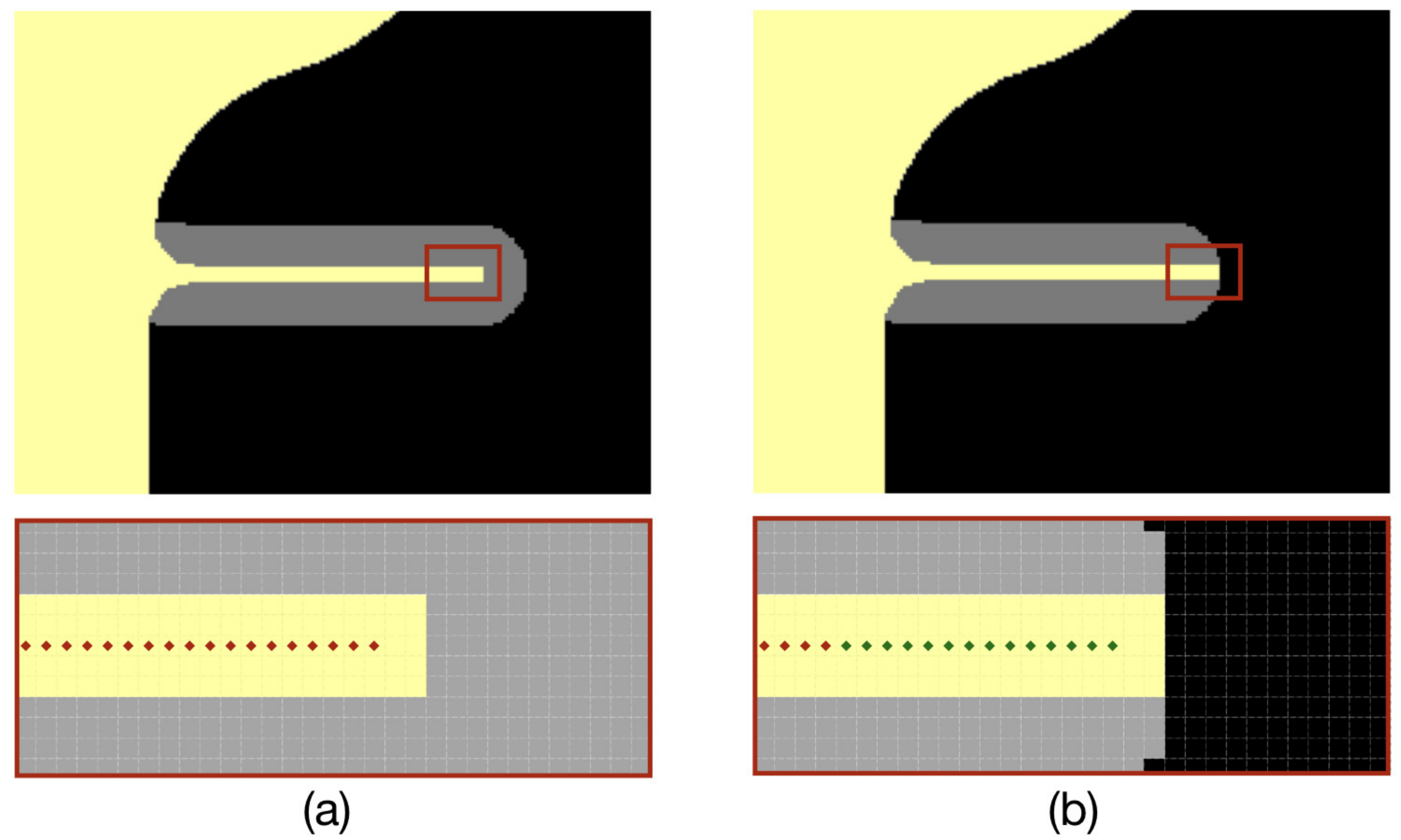}
  \caption{For the same slice (yellow), the protrusion is elongated in the corrected model (gray) when the skeleton is obtained using the thinning process alone (a).  Our spur pixel removal allows us to achieve the intended length by deleting the end-point pixels (marked green) (b). Top row is the original slice (yellow) overlaid with the corrected slice (gray) and the bottom row shows the close-up view of the tip of the protrusion demonstrating the pixel grid. Red marked pixels represent the meso-skeleton pixels.}
  \label{fig:spurPixels}
\end{figure}

\red{Condition $(a)$ and alternating conditions $(c.1)$ and $(c.2)$ ensure that the local connectivity is maintained. Here, the constraint on the crossing number serves as a necessary condition and prevents deletion of pixels in the middle of the medial curves. Condition $(b)$ provides as an endpoint check. While redundant pixels in the middle of the curves are allowed to be deleted, endpoints are detected and preserved through this condition.} With this approach, the object shrinks to an 8-connected skeleton when applied repeatedly, \ie~ an object without holes reduces to a minimally connected stroke whereas an object with holes shrinks to a connected ring halfway between each hole and the outer contour. %Here, the connectedness is ensured through the constraint on the crossing number. 
As the Euler number is preserved, a meso-skeleton $\hat{M}_i$ that is homotopic to $S_i$ is generated. Moreover, this approach allows us to preserve a skeletal structure for geometrically distinct features such as protrusions, in addition to the topologically essential parts (Figure~\ref{fig:distinctFeatures}). For detailed overview of various thinning methodologies and their comparisons to the approach we employed here, the reader is referred to \cite{lam1992thinning}.

For long and thin protrusions, the thinning operation in \eqref{eq:thinning} may converge to the skeletal structure quickly and terminate further removal of pixels. This tends to create undesirable overdepositions later in constructing the corrected model. We demonstrate an example problematic case in Figure~\ref{fig:spurPixels}. As the red marked pixels are assigned to be skeleton pixels in the thinning process, they do not get removed in later iterations. However, following dilation operation of this meso-skeleton with $F$ would extend the protrusion beyond what was originally designed. We address this problem by detecting the spur pixels that are not removed during the thinning process and removing them. To ensure that the protrusion has the intended length in the corrected model, we perform this operation only for the first $d/2$ iterations. A corrected slice obtained with and without spur pixel removal is illustrated in Figure~\ref{fig:spurPixels}.    

\begin{figure}
  \centering
  \includegraphics[width = 0.9\columnwidth]{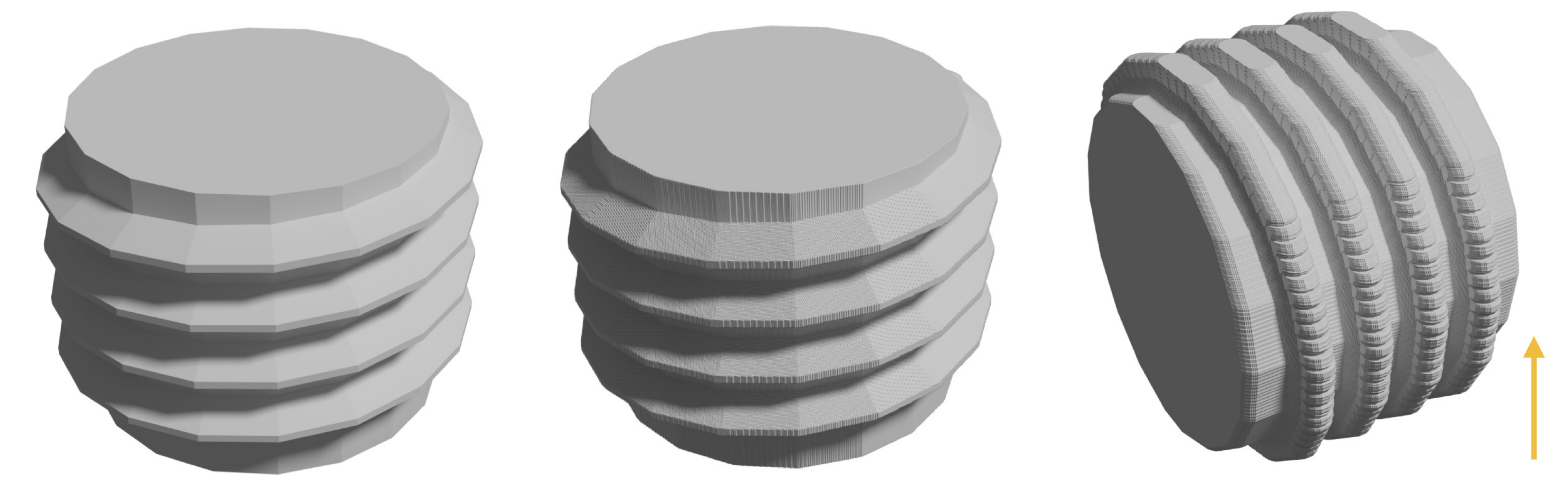}
  \caption{A threaded cylinder (left) can be printed without any corrections in the vertical orientation (middle) whereas a significant amount of corrections are required in the horizontal orientation (right). Build direction is indicated with a yellow arrow.}
  \label{fig:orientationMatters}
\end{figure}

\section{Build Direction Optimization}

As our shape modification method mimics the additive manufacturing process and operates on the slice level, the quality of a printed part and thus, our correction result depends on the selected build direction. The amount of change in the shape (\ie~ amount of added or removed material with the correction process) may be significantly different for different build directions. For instance, a threaded cylinder can be 3D printed in a vertical orientation perfectly, while certain modifications (\eg~ thickening and rounding sharp thread and cylinder edges) are required for the horizontal orientation (Figure~\ref{fig:orientationMatters}) to be printable. Therefore, it is essential to select the right build direction to minimize the difference between the designed shape and the manufactured object. 

%(1) the amount of modification required to print a model and (2) the difference between the manufactured part and the designed model. In our shape modification approach, the former can simply be measured by the amount of overdeposition, \ie~ the amount of additional material required to make the small features manufacturable. As the  

\begin{figure*}
  \centering
  \includegraphics[width = 0.9\textwidth]{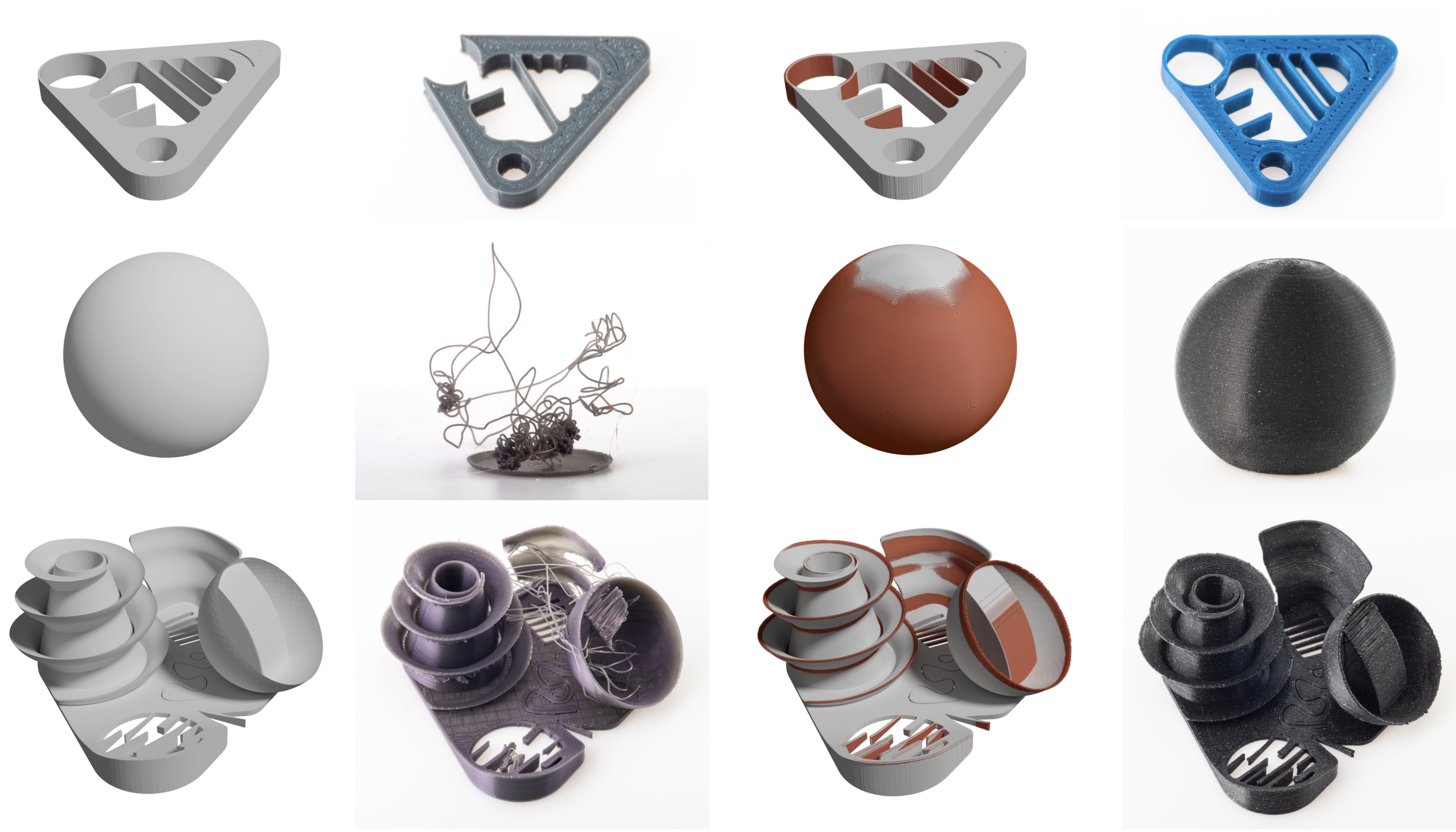}
  \caption{Model correction results. Left-to-right: input uncorrected models with their corresponding failed prints, corrected models and their successful prints. Added and removed material are highlighted in red and blue, respectively. }
  \label{fig:correctionResults}
\end{figure*}

Formal definition of the build direction optimization problem is as follows:

\begin{equation}
\begin{aligned}
& \underset{\boldsymbol{\theta}}{\text{minimize}}
& & C(\boldsymbol{\theta}) = \sum_i | \hat{S}_i(\boldsymbol{\theta}) - S_i(\boldsymbol{\theta}) |   \\
& \text{subject to}
& & \theta_1 \in [-\pi/2, \pi/2] ~ \text{and} ~\theta_2 \in [0, \pi], \\
& & & \boldsymbol{\theta} = [\theta_1, \theta_2]^T. \\
\end{aligned}
\label{Eq:optimizationProblem}
\end{equation}
Here, $\boldsymbol{\theta}$ is the vector of design variables where $\theta_1$ and $\theta_2$ denote the extrinsic Euler angles representing a sequential rotation of the build direction vector about the x and z axes, respectively. Note that the build direction is initialized at +z direction for $\boldsymbol{\theta} = [0,0]^T$. As the slices are identical for any two opposite build directions, we only explore the half-space by setting the bounds for $\theta_1$ and $\theta_2$ as given in \eqref{Eq:optimizationProblem}. 

The objective function in \eqref{Eq:optimizationProblem} measures the absolute difference between the input shape and the corrected shape. Therefore, both the amount of added material through the thickening operations and the amount removed material due to the rounded corners/edges are taken into account in determining the optimum build direction.  Note that it is also possible to enforce the optimizer to favor one over the other (\ie~ material addition might be tolerated in some applications while removal is not) by constructing the objective function as a weighted sum of these two components. Our objective function in \eqref{Eq:optimizationProblem} corresponds to equally weighted sum in such formulation.  

As the algebraic definition of the gradients are not available, we use a derivative-free  simulated annealing approach to solve our optimization problem. We implement the simulated annealing algorithm as described in ~\cite{kirkpatrick1983sa}. The algorithm mimics the thermal annealing process of heating and slowly lowering the temperature of a part to minimize system energy. At each optimization step, we generate new candidate states by sampling a new design vector around the current state. The new states are drawn from a Gaussian distribution. Variance of the distribution is proportional to the temperature of the system. Since the optimization starts with a high temperature value that decreases through the iterations, the extent of the search (\ie ~the distance of the new point from the current point) is higher at the beginning of the optimization allowing a broad search and gets smaller as the optimization converges. Any new state with a lower objective value is accepted. However, the algorithm allows accepting a solution with a higher objective value through a certain probability and alleviates the issue of getting stuck at a local minima. For cooling schedules, we are using exponential cooling in all of our examples. We also experimented with linear and logarithmic cooling schedules and observed similar convergence properties.

\begin{figure}
  \centering
  \includegraphics[width = \columnwidth]{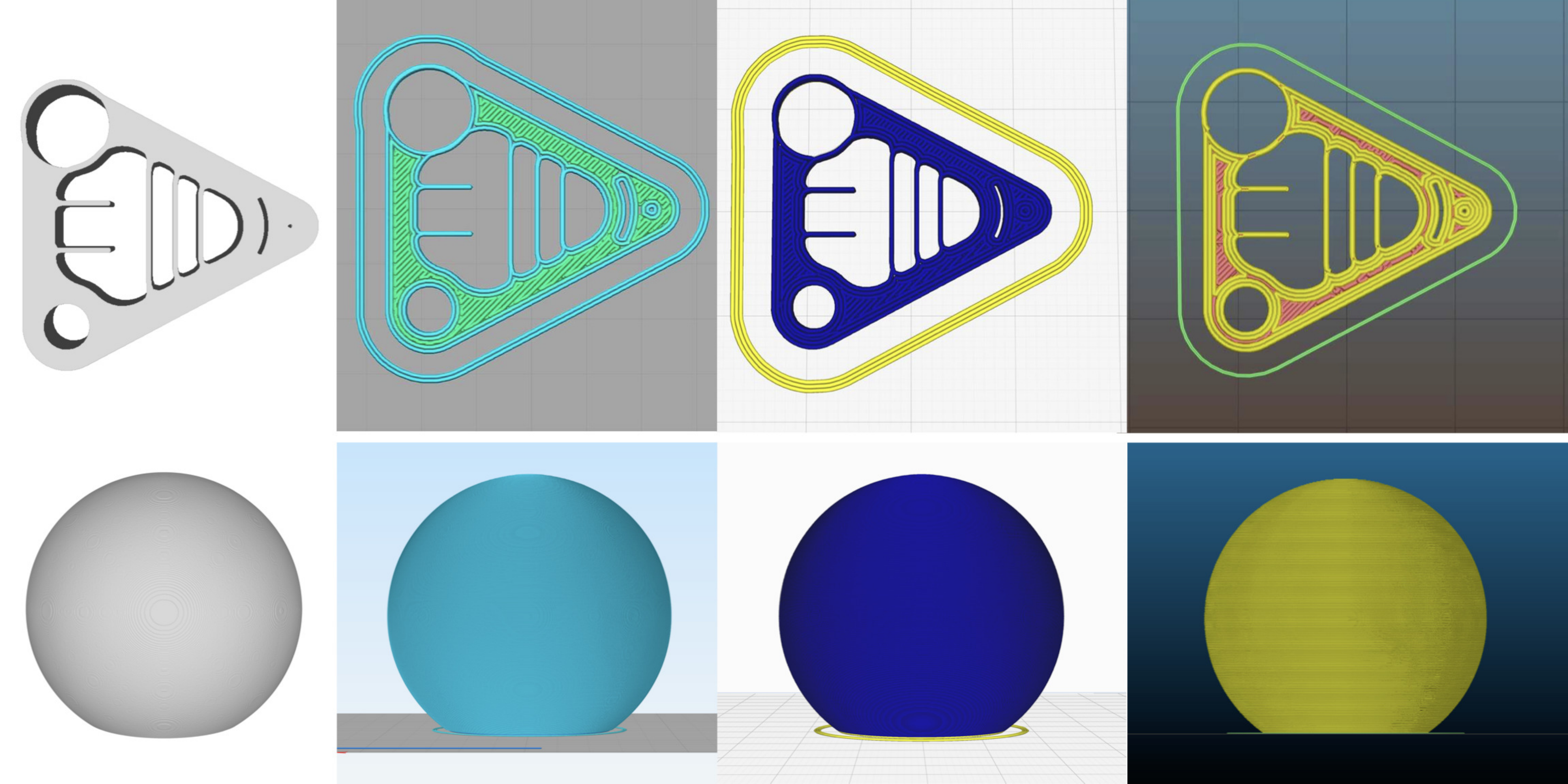}
  \caption{For the corrected models, slicer software can generate toolpaths successfully without creating any ambiguities. Left-to-right: the corrected model, toolpaths generated by Simplify3D \cite{simplify3d}, Ultimaker Cura \cite{cura} and Slic3r (Prusa edition) \cite{slic3r}.}
  \label{fig:noAmbiguities}
\end{figure}

\section{Results and Discussion}
\paragraph{Model Correction}
Figure~\ref{fig:correctionResults} illustrates the results of our model correction on various models. Here, we assume the build direction to be $\boldsymbol{\theta} = [0,0]$. Our meso-skeleton approach detects the problematic regions and modifies each layer automatically to make them manufacturable. Printed results verify our method that manufacturing failures are resolved in the corrected models. Note that in some extreme cases such as the uniform thickness spherical shell model (middle row), our correction makes the completely unprintable model printable using the same manufacturing process and same set of process parameters. Similarly, disappearing regions that are geometrically or topologically important in the design is recovered for the test part (top row) and the playground (bottom row). Table~\ref{tab:correctionResults} summarizes our model correction results with various metrics relevant to these models.

\begin{table}
\setstretch{1.0}
\small
\caption{Performance of our model correction algorithm on a variety of models. } 
\centering % centering table
\begin{tabular}{l ccc} % creating columns
& & & \\ % put some space after the caption
\hline
Model & Slices & Image Size & Time [s] \\ [0.5ex]
\hline
Test part							& 	65 & [583 $\times$ 597]	& 3.44\\
Molecule							& 	383 & [513 $\times$ 462]	& 11.22\\
Threaded cylinder				& 	421 & [623 $\times$ 632]	& 44.78\\
Spherical	shell						& 	712 & [795 $\times$ 795]	& 67.96\\
Playground						& 	555 & [1417 $\times$ 1479]	& 187.57\\
Dragon						& 	687 & [284 $\times$ 253]	& 4.86\\
Helical tower						& 	1472 & [376 $\times$ 382]	& 27.75\\
\hline
\end{tabular}
\label{tab:correctionResults}
\end{table}

\begin{figure}
  \centering
  \includegraphics[width = \columnwidth]{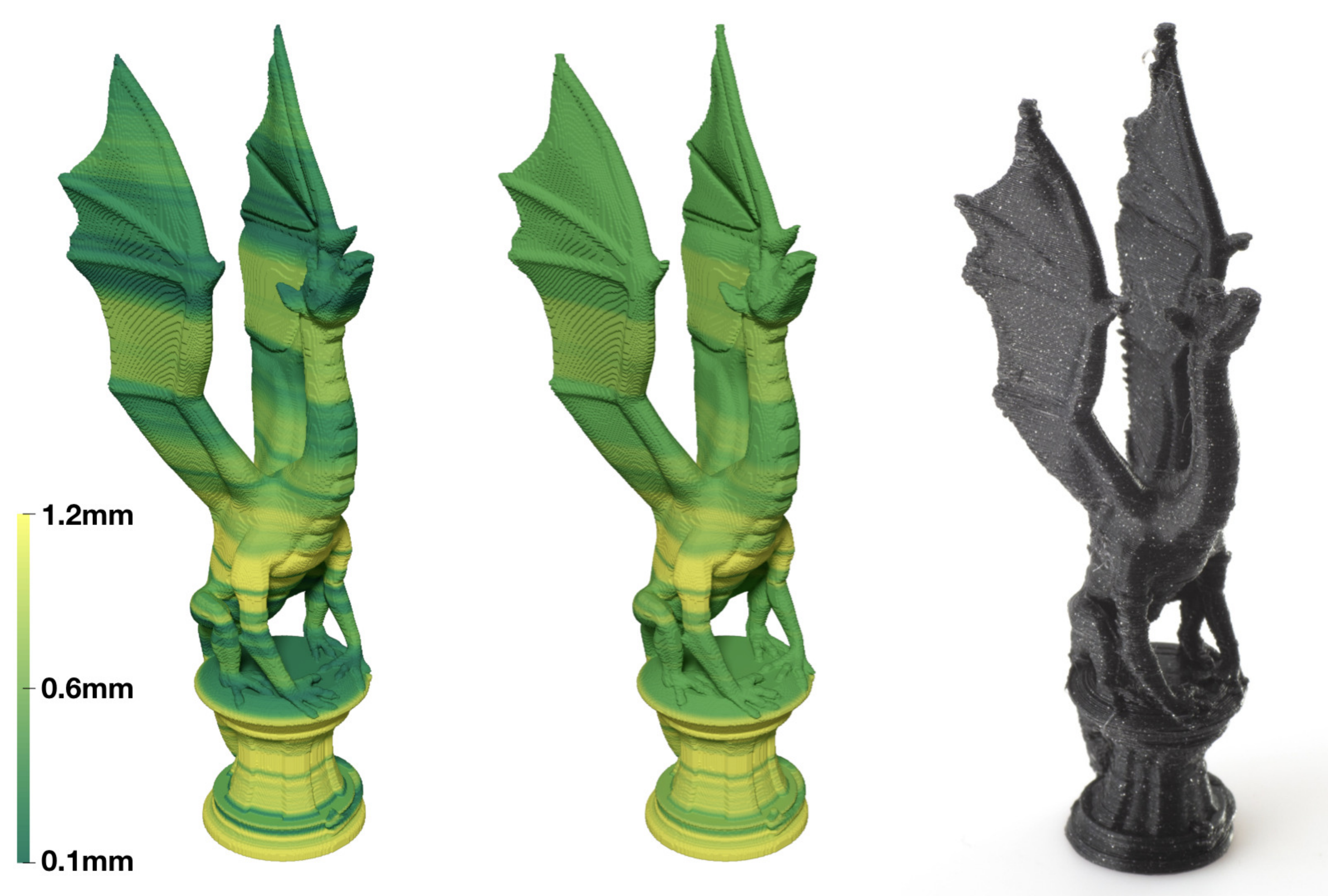}
  \caption{Distribution of the maximum allowable structuring element size suggested by our algorithm. Left-to-right: corrected models obtained using varying structuring element size range of $[0.1,1.2]mm$, $[0.6,1.2]mm$ and the corresponding printed model.}
  \label{fig:varyingExtrusionWidth}
\end{figure}

In Figure~\ref{fig:noAmbiguities}, we show the toolpaths generated by different slicer software for our corrected test part and spherical shell models. Unlike the toolpaths generated for the original models (Figure~\ref{fig:slicerExamples}), there are no ambiguities in the resulting shape, \ie~ all the slicers were able to generate similar results with no indication of any fabrication failure.

\paragraph{Structuring Element Size}
As the meso-skeleton is generated through iterative removal of contour pixels, we can compute the exact state where the skeleton is reduced to a single pixel wide line for the first time. At this state, the number of removed contour layers corresponds to the maximum allowable structuring element size to print the current slice without making any topologically or geometrically important changes. This allows us to suggest a per-layer structuring element size that can be used to fabricate the input geometry without requiring any correction. Figure~\ref{fig:varyingExtrusionWidth} illustrates the distribution of the required structuring element size on an example model. It can be observed that for layers involving thin features (such as the tip of the wings, the claws and the wing membranes), our algorithm suggests smaller structuring elements as expected. In order to validate this result, we have modified an open-source slicer (gsSlicer \cite{gsSlicer}) to adjust the extrusion width for each layer individually in the FDM toolpath generation. Due to the physical constraints of the printer, we limited the change in the extrusion width to $2\times$ only by setting the allowable range to $0.6-1.2mm$. In that case, it can be observed in Figure~\ref{fig:varyingExtrusionWidth} (middle) that layers containing critical features are assigned the lower bound value. Such an adjustment in the extrusion width results in $\sim8\%$ decrease in the build time as the layers without any critical features can be printed faster using a larger extrusion width. The improvement in build time can be more significant for complex models containing large bulky regions with more localized small and thin features.

\begin{figure}
  \centering
  \includegraphics[width = \columnwidth]{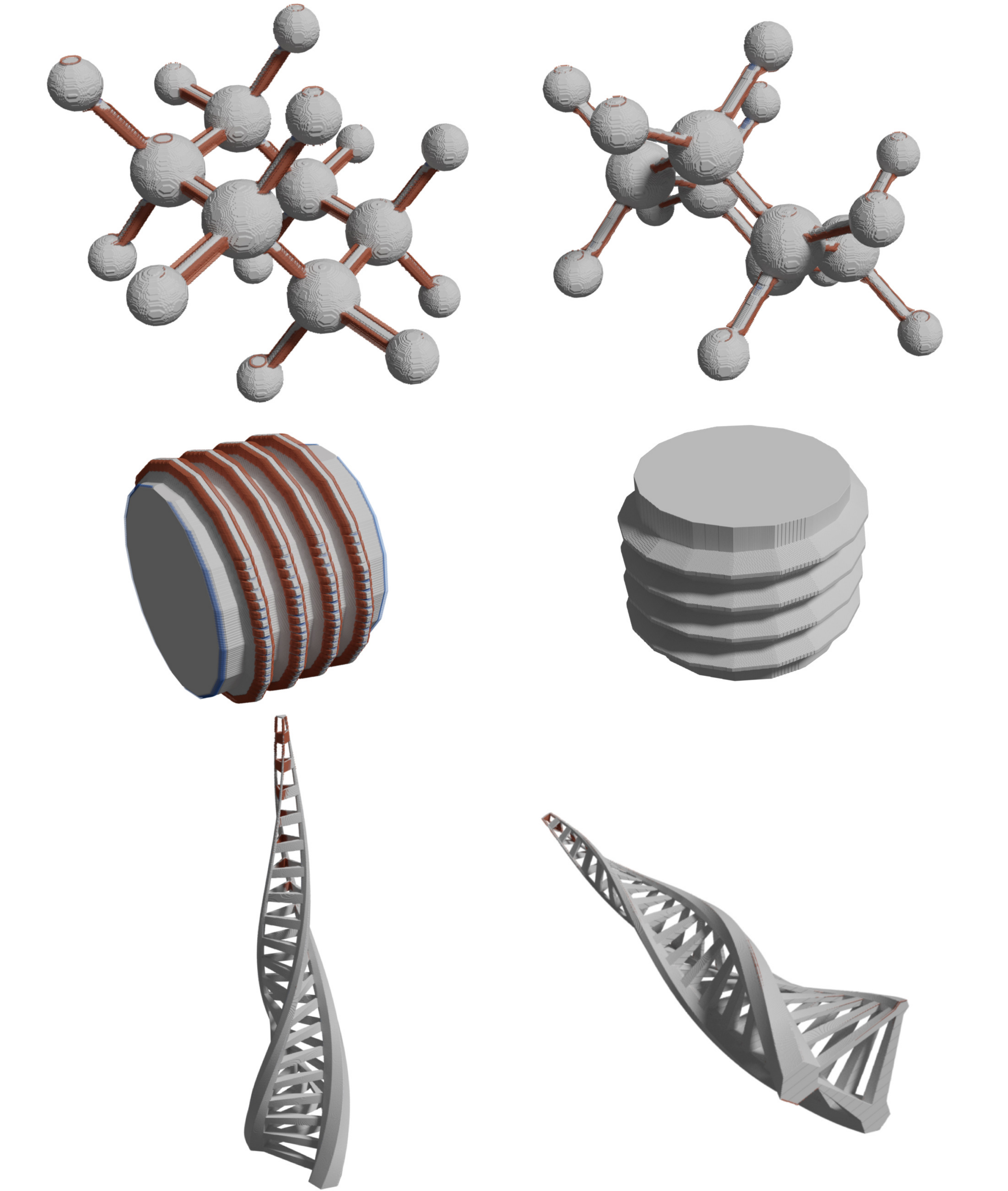}
  \caption{Build direction optimization results. The amount of corrections required to make the model manufacturable is significantly smaller in the optimized orientation (right) compared to the arbitrary initial orientation (left). Added and removed material are highlighted in red and blue, respectively. }
  \label{fig:orientationOptimizationResults}
\end{figure}

\begin{table*}
\setstretch{1.0}
\small
\caption{Performance of our build direction optimization on a variety of models. } 
\centering % centering table
\begin{tabular}{l cccc} % creating columns
& & & & \\ % put some space after the caption
\hline
\multirow{2}{*}{Model} & \multicolumn{2}{c}{Objective Value} & \multirow{2}{*}{$\boldsymbol{\theta}$} &\multirow{2}{*}{Time} \\ 
\cline{2-3}
& Initial & Optimized & \\[0.5ex]
\hline
Molecule							& 	180661 & 118927 & [-0.75, 0.03] & 62m\\
Threaded cylinder				& 	2040590 & 26687	& [1.52, 0.00] & 3h 56m\\
Dragon				& 	131748 & 63694 & [-1.98, 1.57] & 35m \\
Helical tower				& 	134276 & 112601 & [-1.51, 1.09] & 2h 31m \\
\hline
\end{tabular}
\label{tab:orientationOptimizationResults}
\end{table*}

\paragraph{Build Direction}
Figure~\ref{fig:orientationOptimizationResults} and Figure~\ref{fig:teaser} illustrate our build direction optimization results. We achieved $\sim15\%$ to $\sim50\%$ reduction in the objective value for the molecule,  the dragon and helical tower models. In order to validate our optimization approach, we also experimented with the threaded cylinder model. Our optimizer was able to converge to the obvious solution of the vertical print configuration at which the model can be printed almost perfectly. It can be observed that major problems in the horizontal orientation including the rounding of the cylinder edges and the thickening of the sharp threads are avoided in the vertical orientation, resulting in $\sim98\%$ reduction in the objective value. Table~\ref{tab:orientationOptimizationResults} summarizes our build direction optimization results together with other performance metrics.

\subsection{Validation and Performance}
\paragraph{Fabrication}
We 3D printed the original designs as well as the corrected models using a consumer level FDM printer with PLA material. We use Simplify3D \cite{simplify3d} to generate the toolpaths for our prints. In order to compare the results, we have kept all the printer and slicer parameters the same for both the original designs and the corrected models. Although we have only experimented with FDM printing technology, our approach is applicable to a wide variety of additive manufacturing methods where the printing process is driven by the planar translation of a material deposition or fusion mechanism, including but not limited to stereolithography, sintering and binding based methods.

\paragraph{Performance}
We tested our method on a 3.7GHz 8 Core Intel Xeon computer with 32GB of memory. As each slice can be processed independently, we use parallel processing in our shape modification algorithm. Our parallel implementation is only limited the outer-most loop where the individual slices are processed. However, the thinning algorithm can also be parallelized for further improvement in the performance. We use OpenVDB~\cite{museth2013openvdb} for sparse voxel operations.

Table~\ref{tab:correctionResults} shows the performance of our correction algorithm for a predetermined build direction ($\boldsymbol{\theta} = [0,0]$). As the thinning operation is performed repeatedly for each slice, it constitutes the majority of the computational cost in our approach. Therefore, the performance is heavily affected by both the number of slices as well as the size of each slice (\ie~ image size). 

Table~\ref{tab:orientationOptimizationResults} reports the performance of our build direction optimization algorithm. In all of our examples, the optimization converges under 300 iterations. As the number of slices and the image size change for each new candidate build direction, average per iteration processing time may change significantly in comparison to the single iteration time reported in Table~\ref{tab:correctionResults}. We have observed $10\%$ to $40\%$ increase in our example models. As only the per-slice operations are parallelized in our implementation, we observe longer processing times for candidate orientations that have smaller number of slices with larger image sizes.

\subsection{Limitations and Future Work}
In our model correction approach, we depend on the meso-skeleton for topological correctness of the resulting model. We found this approach to work well in preserving the topology for models where the topologically important thin features are not very close to each other. However, for shapes where the minimum distance between two thin features are smaller than the structuring element diameter, the topology may change when the meso-skeleton is dilated with the structuring element. \red{For example, when there is a small channel or tunnel close to a thin feature that is required to be corrected, it may be filled and disappear due to the thickening operation on the thin feature next to it.} Although detection of these local topological changes is possible \cite{behandish2019detection}, model correction under such problematic cases remains an open problem. 

\red{Our approach relies on voxel/pixel based representation of the model and individual slices in various stages of the correction and the optimization processes. Although it is a widely accepted practical approach in discretizing the 3D shape, it introduces larger aliasing error compared to the other boundary conformal representations including ray-rep and its variations \cite{chen2013regulating}. Complementing our approach with such representations could improve the quality of the resulting models and corresponding printed parts.}

As the meso-skeleton created for thin features is approximately in the middle with respect to the contour, the resulting correction thickens the object equally in both directions. However, in certain applications, this may not be desirable due to the geometrical constraints. A natural extension of our approach would be to allow unidirectional corrections such that the geometrical tolerances or constraints are satisfied.  

In our build direction optimization, we only focus on minimizing the change in the shape. However, build direction affects many other properties significantly, including support structures, strength and surface roughness. Such properties may also influence the printability of a model. In future, our optimization could be extended to take these effects into account and select a build direction minimizing the negative aspects concurrently to improve the printability.

\section{Conclusion}
We present a model correction and build orientation optimization method for manufacturability. 
We propose a thinning based approach to compute the allowable space a print head can traverse in each slice during the printing process. With this approach, we show that both topologically and geometrically important features that are smaller than the minimum printable feature size are thickened automatically to allow a minimal path for the print head to deposit material. Combined with our build direction optimization, we demonstrate that the printability of a model can be improved with a minimal amount of shape modifications. We evaluate the performance of our method on variety of 3D models. Our results show that corrected model accurately approximates the printed part reducing ambiguities in the downstream design-to-manufacturing pipeline.

\begin{acknowledgment}
The authors would like to thank Chaman Singh Verma for his help with OpenVDB. We are grateful to the designers whose 3D models we used: Louise Driggers (Thingiverse) for dragon, marcaxi (Thingiverse) for threaded cylinder and Daniel Lu (Grabcad) for molecule. This research was developed with funding from the Defense Advanced Research Projects Agency (DARPA), United States. The views, opinions and/or findings expressed are those of the authors and should not be interpreted as representing the official views or policies of the Department of Defense or U.S. Government.
\end{acknowledgment}

%%%%%%%%%%%%%%%%%%%%%%%%%%%%%%%%%%%%%%%%%%%%%%%%%%%%%%%%%%%%%%%%%%%%%%
% The bibliography is stored in an external database file
% in the BibTeX format (file_name.bib).  The bibliography is
% created by the following command and it will appear in this
% position in the document. You may, of course, create your
% own bibliography by using thebibliography environment as in
%
% \begin{thebibliography}{12}
% ...
% \bibitem{itemreference} D. E. Knudsen.
% {\em 1966 World Bnus Almanac.}
% {Permafrost Press, Novosibirsk.}
% ...
% \end{thebibliography}

% Here's where you specify the bibliography style file.
% The full file name for the bibliography style file 
% used for an ASME paper is asmems4.bst.
\bibliographystyle{asmems4}

% Here's where you specify the bibliography database file.
% The full file name of the bibliography database for this
% article is asme2e.bib. The name for your database is up
% to you.
\bibliography{asme2e}

\end{document}